\documentclass[twocolumn,showpacs,preprintnumbers,amsmath,amssymb,showkeys,nofootinbib]{revtex4}

\usepackage{graphicx}
\usepackage{dcolumn}
\usepackage{bm}
\newcommand{\bqa}{\begin{eqnarray}}
\newcommand{\eqa}{\end{eqnarray}}
\newcommand{\be}{\begin{equation}}
\newcommand{\ee}{\end{equation}}

\begin{document}


\title{Peak shifts due to $B^{(*)}-\bar{B}^{(*)}$ rescattering in $\Upsilon(5S)$ dipion transitions }

\author{Ce Meng$~^{(a)}$ and Kuang-Ta Chao$~^{(a,b)}$}
\affiliation{ {\footnotesize (a)~Department of Physics and State Key
Laboratory of Nuclear Physics and Technology, Peking University,
 Beijing 100871, China}\\
{\footnotesize (b)~Center for High Energy Physics, Peking
University, Beijing 100871, China}}


\begin{abstract}

We study the energy distributions of dipion transitions
$\Upsilon(5S)$ to $\Upsilon(1S,2S,3S)\pi^+\pi^-$ in the final state
rescattering model. Since the $\Upsilon(5S)$ is well above the open
bottom thresholds, the dipion transitions are expected to mainly
proceed through the real processes $\Upsilon(5S)\to
B^{(*)}\bar{B}^{(*)}$ and $B^{(*)}\bar{B}^{(*)}\to
\Upsilon(1S,2S,3S)\pi^+\pi^-$. We find that the energy distributions
of $\Upsilon(1S,2S,3S)\pi^+\pi^-$ markedly differ from that of
$\Upsilon(5S)\to B^{(*)}\bar{B}^{(*)}$. In particular, the resonance
peak will be pushed up by about 7-20 MeV for these dipion
transitions relative to the main hadronic decay modes. These
predictions can be used to test the final state rescattering
mechanism in hadronic transitions for heavy quarkonia above the open
flavor thresholds.
\end{abstract}


\maketitle

\section{Introduction}

Hadronic transitions between heavy quarkonia are important for
understanding both the heavy quarkonium dynamics and the formation
of light hadrons (for recent reviews see, e.g.~\cite{Kuang06}).
Particularly, in recent years, the dipion and single pion
transitions have proved to be a very efficient way to discover new
or missing charmonium or charmonium-like states, the so-called
"$X,Y,Z$" mesons, by Belle, Babar, and CLEO collaborations
(see~\cite{Olsen} for a recent review and related references). To
search for the partners of $X$ and $Y$ in the $b\bar{b}$ sector, say
$X_b$ and $Y_b$, a similar approach was also suggested~\cite{hou}.
More recently, Belle has found a very striking
result~\cite{Belle07-5SmS} that signals of $\Upsilon(mS)\pi^+\pi^-$
with $m=1,2,3$ collected at the energy of 10870 MeV, i.e. at the
peak of $\Upsilon(5S)$, indicate that the partial widths of
$\Upsilon(5S)\to\Upsilon(mS)\pi^+\pi^-$ are larger than that of
$\Upsilon(4S)\to\Upsilon(1S,2S)\pi^+\pi^-$ by two orders of
magnitude or more if the $\Upsilon(5S)$ is the sole source of the
observed dipion transition events.

If the peak at $\Upsilon(10870)$ is identified with the
$\Upsilon(5S)$ resonance, we will have to answer the question: what
is the essential difference between the $\Upsilon(5S)$ and
$\Upsilon(4S)$ in their dipion transitions.  To answer this
question, in an earlier paper~\cite{Meng-Y5S} we use the final state
rescattering model~\cite{Cheng05_ReSC} to study the dipion
transitions of $\Upsilon(5S)$ and $\Upsilon(4S)$. In this model, the
$\Upsilon(5S/4S)$ first decays to $B^{(*)}\bar{B}^{(*)}$, and then
the $B$ meson pair turns into a lower $\Upsilon$ state and two pions
through exchange of another $B^{(*)}$ meson. Since the real
rescattering contributions are expected to be dominant in these
processes, the difference between $\Upsilon(5S)$ and $\Upsilon(4S)$
dipion transitions can be explained mainly by the difference in
available phase space for the $\Upsilon(5S/4S)\to
B^{(*)}\bar{B}^{(*)}$ decays. Although sharing the same quantum
numbers and having similar leptonic widths, the $\Upsilon(5S)$ and
$\Upsilon(4S)$ are dramatically different in their hadronic
transitions.  In the real rescattering process where the $b\bar b$
resonance is above the open bottom threshold, the amplitude is
basically proportional to the probability of the corresponding open
bottom decay, which is proportional to the P-wave phase space factor
$|\vec{p}_1|^3$~\cite{Meng-Y5S,Simonov07-nSmS}. Here, $\vec{p}_1$
denotes the 3-momentum of $B^{(*)}$ or $\bar{B}^{(*)}$ in the rest
frame of $\Upsilon(5S/4S)$. Note that the $p$-value $|\vec{p}_1|$ of
the decay $\Upsilon(5S)\to B\bar{B}$ is about 3.84 times larger than
that of $\Upsilon(4S)\to B\bar{B}$, and this fact will mainly result
in a huge difference, which is about a factor of 200-600 in
magnitude~\cite{Meng-Y5S}, between the partial widths of dipion
transitions of $\Upsilon(5S)$ and $\Upsilon(4S)$. (Note that the
$|\vec{p}_1|^6$ factor would give an enhancement of about 3200 for
the $\Upsilon(5S)$ relative to $\Upsilon(4S)$ but an effective form
factor associated with the coupling constants will lower the
enhancement factor by about an order of magnitude, see discussions
in next sections.) As a result of the rescattering mechanism, it
might be unnecessary to introduce an exotic interpretation of
$\Upsilon(5S)$ resonance or a $Y_b$ state to account for the
experimental data~\cite{Belle07-5SmS}.

To further clarify the issue mentioned above, it is useful to study
other features of the final state rescattering mechanism. In this
paper we will show that a distinct and important consequence of the
final state rescattering mechanism is the peak shift effect, which
is related to the unique energy dependence of the cross section of
$e^+e^-\to\Upsilon(5S)\to \Upsilon(mS)\pi^+\pi^-$.  As a result of
the final state rescattering process
$B^{(*)}\bar{B}^{(*)}\to\Upsilon(mS)\pi^+\pi^-$, the energy
distributions of the $\Upsilon(5S)\to\Upsilon(mS)\pi^+\pi^-$ events
will differ from that of $\Upsilon(5S)\to B^{(*)}\bar{B}^{(*)}$. In
particular, the observed resonance peak in
$\Upsilon(5S)\to\Upsilon(mS)\pi^+\pi^-$ is expected to be pushed up
to higher energies markedly compared with that of $\Upsilon(5S)\to
B^{(*)}\bar{B}^{(*)}$. The physical reason for this peak shift is
quite evident that at higher energies around the $\Upsilon(5S)$
resonance peak the rescattering process will acquire more phase
space and then get a larger rate. This peak shift effect is an
inevitable result of the final state rescattering, and therefore it
can serve as a crucial test for this mechanism. In the following, we
will study this effect in a more quantitative way.

\section{The rescattering model}

\begin{figure}[t]
\begin{center}
\vspace{0cm}
 \hspace*{0cm}
\scalebox{0.5}{\includegraphics[width=16cm,height=13cm]{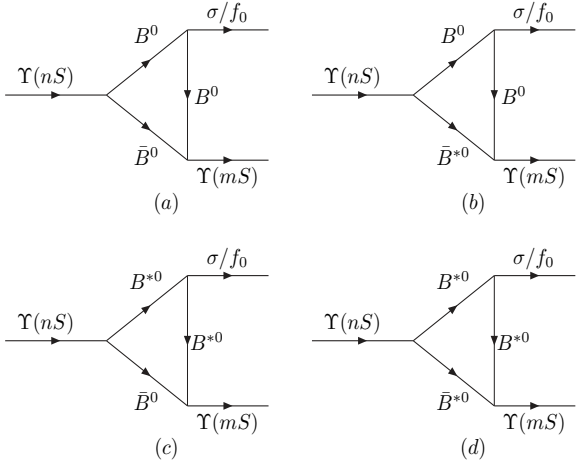}}
\end{center}
\vspace{0cm}\caption{Typical diagrams for $\Upsilon(nS)\to
B^{(*)}\bar{B}^{(*)}\to\Upsilon(mS)\mathcal{S}$. Other diagrams can
be obtained by charge conjugation transformation
$B\leftrightarrow\bar{B}$ and isospin transformation
$B^0\leftrightarrow B^+$ and $\bar{B}^0\leftrightarrow
B^-$.}\label{Fig:Y-YS}
\end{figure}

As in Ref.~\cite{Meng-Y5S}, we assume that in the $\Upsilon(4S,5S)$
dipion transitions the two pions are produced mainly via scalar
resonances coupled  to intermediate $B^{(*)}$ mesons due to the
long-distance final state interactions. The typical rescattering
diagrams for $\Upsilon(4S,5S)\to\Upsilon(1S,2S)\mathcal{S}$ are
shown in Fig.~\ref{Fig:Y-YS}, and the others can be related to those
in Fig.~\ref{Fig:Y-YS} by charge conjugation transformation
$B\leftrightarrow\bar{B}$ and isospin transformation
$B^0\leftrightarrow B^+$ and $\bar{B}^0\leftrightarrow B^-$. Here,
$\mathcal{S}$ denotes scalar resonance $\sigma$ or $f_0(980)$
(perhaps also $f_0(1370)$), which will decay to $\pi\pi(K\bar{K})$
eventually.

In Fig.~\ref{Fig:Y-YS}, the intermediate states
$B^{(*)}\bar{B}^{(*)}$ can be real or virtual, which corresponds
respectively to the imaginary part or the real part of the
amplitude. In Ref.~\cite{Meng-Y5S}, we argued that in general it is
the real rescattering process that is dominant unless the resonance
is very close to the open flavor threshold.
More quantitatively, we used the dispersion relation to estimate the
virtual rescattering contributions and found that they are extremely
small for $\Upsilon(5S)$ because  $\Upsilon(5S)$ is far above the
$B\bar B$ threshold, whereas for $\Upsilon(4S)$, they can be
comparable to the real ones but with large
uncertainties~\cite{Meng-Y5S}. Nevertheless, this does not affect
the calculated large difference between the transition widths of
$\Upsilon(5S)$ and $\Upsilon(4S)$. (In contrast, in the case of
$X(3872)$~\cite{Meng07_X3872_ReSC} or
$Z(4430)$~\cite{Meng07_Z4430_ReSC}, the virtual rescattering effects
could be dominant due to the extreme closeness between the resonance
mass and the open flavor threshold.)

In the real rescattering processes, the amplitudes of
Fig.~\ref{Fig:Y-YS}(a,b,c,d) are dominated by their
absorptive(imaginary) parts, which can be derived by the Cutkosky
rule as~\cite{Meng-Y5S}:
\bqa\label{Abs:CutRule}
\textbf{Abs}_i&=&\frac{|\vec{p}_1|}{32\pi^2m_{\Upsilon(nS)}}\int
d\Omega
\mathcal{A}_i(\Upsilon(nS)\to B^{(*)}\bar{B}^{(*)})\nonumber\\
&&\times \mathcal{C}_i(B^{(*)}\bar{B}^{(*)}\to
\Upsilon(mS)\mathcal{S}), \eqa
where $i=(a,b,c,d)$, and $d\Omega$ and $\vec{p}_1$ denote the solid
angle of the on-shell $B^{(*)}\bar{B}^{(*)}$ system and the
3-momentum of the on-shell $B^{(*)}$ meson in the rest frame of
$\Upsilon(nS)$, respectively.

The amplitudes $\mathcal{A}_i$ and $\mathcal{C}_i$ are determined by
the effective Lagrangians~\cite{Meng-Y5S}
\begin{subequations} \label{effective-Lagrangians}
\begin{eqnarray}
\mathcal{L}_{\Upsilon BB}&=& g_{\Upsilon
BB}\Upsilon_\mu(\partial^\mu
B{B}^{\dagger}-B\partial^\mu {B}^{\dagger}),\label{L-YBB}\\
\mathcal{L}_{\Upsilon B^*B}&=& \!\frac{g_{\Upsilon\!
B^*\!B}}{m_{\Upsilon}}\varepsilon^{\mu\nu\alpha\beta}\partial_\mu
\!\Upsilon_\nu
\!\nonumber\\
&& \times(B^*_\alpha\overleftrightarrow{\partial}_\beta
{B}^{\dagger}\!\! - \!\!
B\overleftrightarrow{\partial}_\beta{B}^{*\dagger}_\alpha\!),\label{L-YB*B}\\
\mathcal{L}_{\Upsilon B^*B^*}&=& g_{\Upsilon B^* B^*} (
-\Upsilon^\mu
B^{*\nu}\overleftrightarrow{\partial}_\mu {B}_\nu^{*\dagger} \nonumber\\
&&+ \Upsilon^\mu B^{*\nu}\partial_\nu{B}^{*\dagger}_{\mu} -
\Upsilon_\mu\partial_\nu B^{*\mu}
{B}^{*\nu\dagger}),\label{L-YB*B*}\\
\mathcal{L}_{\mathcal{S} BB}&=& g_{\mathcal{S} BB}\mathcal{S}
B{B}^{\dagger},\label{L-SBB}\\
\mathcal{L}_{\mathcal{S} B^*B^*}&=& -g_{\mathcal{S}
B^*B^*}\mathcal{S} B^*\cdot{B^*}^{\dagger},\label{L-SB*B*}
\end{eqnarray}
\end{subequations}
where
$\overleftrightarrow{\partial}=\overrightarrow{\partial}-\overleftarrow{\partial}$.
Here, the heavy quark symmetry and chiral symmetry in
(\ref{effective-Lagrangians}) are basically ensured by the kinematic
conditions.

Following Ref.~\cite{Meng-Y5S}, we choose the on-shell coupling
constants
\bqa g_{\Upsilon(mS)B^{(*)}B^{(*)}}&=& 24,~~~~m\leq4\label{g:Upsilon(4S)BB}\\
 g_{\Upsilon(5S)BB}  &=& 2.5,\label{g:Upsilon(5S)BB}\\
 g_{\Upsilon(5S)B^*B}&=& 1.4\pm0.3,\label{g:Upsilon(5S)B*B}\\
 g_{\Upsilon(5S)B^*B^*}&=& 2.5\pm0.4,\label{g:Upsilon(5S)B*B*}\eqa
where the values in (\ref{g:Upsilon(4S)BB}) and
(\ref{g:Upsilon(5S)BB}-\ref{g:Upsilon(5S)B*B*}) are determined by
the measured widths $\Gamma(\Upsilon(4S)\to B\bar{B})$ and
$\Gamma(\Upsilon(5S)\to B^{(*)}\bar{B}^{(*)})$, respectively. On the
other hand, the coupling constants $g_{\mathcal{S}B^{(*)}B^{(*)}}$
can be related to the well-known one
$g_{D^*D\pi}$~\cite{Meng07_X3872_ReSC} by heavy quark flavor
symmetry and chiral symmetry, and we choose~\cite{Meng-Y5S}
\bqa g_{\sigma BB}&=& g_{\sigma B^*B^*}= 10~\mbox{GeV},\nonumber\\
 g_{f0 BB}&=& g_{f0 B^*B^*}= 10\sqrt{2}~\mbox{GeV}.\nonumber\eqa

To account for the off-shell effect of the exchanged $B^{(*)}$ meson
in Fig.~\ref{Fig:Y-YS}, one need introduce form factors, such
as~\cite{Cheng05_ReSC}
\be \mathcal{F}_1(m_{i},q^2)=\frac{(\Lambda+m_i)^{2}-m_{i}^2
}{(\Lambda+m_i)^{2}-q^{2}},\label{formfactor1} \ee
to the vertexes $\mathcal{S}BB$ and $\Upsilon(mS)BB$. We will fix
the cutoff $\Lambda=660$ MeV~\cite{Meng-Y5S} in our numerical
analysis in the next section.

We treat the scalar resonance $\mathcal{S}$ as a narrow one and use
the Breit-Wigner distribution
\be \mathcal{F}_{\mathcal{S}}(t)=\frac{1}{\pi}\frac{\sqrt
t\Gamma_{\mathcal{S}}(t)}{(t-m_{\mathcal{S}}^2)^2+m_{\mathcal{S}}^2\Gamma_{\mathcal{S}}(t)^2}\label{BW:scalar}\ee
to describe the resonance in the calculation of cross sections, as
the treatment of $\rho$ resonance in Ref.~\cite{Meng07_X3872_ReSC}.
In (\ref{BW:scalar}), the variable $t$ denotes the momentum squared
of $\mathcal{S}$, and the function $\Gamma_{\mathcal{S}}(t)$ is
given by
\bqa
&&\Gamma_{\mathcal{S}}(t)=\frac{p_{\pi}g_{\mathcal{S}\pi\pi}}{8\pi
t}+\frac{p_{K}g_{\mathcal{S}KK}}{8\pi t},\label{Total-width:scalar}\\
&&p_{\pi}=\sqrt{\frac{t}{4}-m_{\pi}^2},~~~p_{K}=\sqrt{\frac{t}{4}-m_{K}^2}\nonumber\eqa
The resonance parameters in (\ref{BW:scalar}) and the coupling
constants in (\ref{Total-width:scalar}), which are listed in
Tab.~\ref{Tab:Scalar-resonance}, are chosen mostly from
Ref.~\cite{Komada01-scalar-contributions}, except for $m_{f_0(980)}$
from PDG2006~\cite{PDG06}.

Although the Breit-Wigner description (\ref{BW:scalar}) for the
scalar resonance is somehow rough, especially for the $\sigma$, it
is efficient enough to get the order of magnitude of the widths
$\Gamma(\Upsilon(4S/5S)\to\Upsilon(mS)\pi\pi)$
correctly~\cite{Meng-Y5S}. Moreover, since the final state phase
space (FSPS) of $\Upsilon(5S)\to\Upsilon(mS)\mathcal{S}$ is smeared
by the large width of $\mathcal{S}$, one would not expect that this
FSPS could bring on large energy dependence of the cross section,
unless it happens to be near the threshold, such as the case of
$\Upsilon(5S)\to\Upsilon(3S)\sigma$. We will discuss this further in
the next section.

\begin{center}\begin{table}
\caption{Resonance parameters of $\sigma$ and
$f_0(980)$~\cite{Komada01-scalar-contributions,PDG06}.}

\begin{tabular}{cccccc}
\hline\hline
   & $m_\mathcal{S}$ & $g_{\mathcal{S}\pi\pi}$ & $\Gamma_{\mathcal{S}\pi\pi}$ & $g_{\mathcal{S} KK}$ & $\Gamma_{\mathcal{S} KK}$  \\
   & (MeV) & (GeV) & (MeV) & (GeV) & (MeV)  \\
\hline
$\sigma$   & $526\pm30$ & 3.06 & $302\pm10$ &       &          \\
$f_0(980)$ & $980\pm10$ & 1.77 & $61\pm1$   & 2.70  & $12\pm1$ \\
\hline\hline \label{Tab:Scalar-resonance}\end{tabular}

\end{table}\end{center}

\section{Peak shifts in $\Upsilon(5S)$ dipion transitions}

In the absorptive part $\textbf{Abs}_i$ in (\ref{Abs:CutRule}), the
amplitude $\mathcal{A}_i$ is proportional to $|\vec{p}_1|$ since it
involves an on-shell P-wave vertex $\Upsilon(nS) B^{(*)}B^{(*)}$.
Furthermore, a hidden factor $|\vec{p}_1|$ will emerge after
performing the integral in (\ref{Abs:CutRule}) explicitly. As a
result, the amplitude $\textbf{Abs}_i$ is proportional to
$|\vec{p}_1|^3$. As we have mentioned above, this brings on the
strong $s$-dependence of the cross section of
$e^+e^-\to\Upsilon(5S)\to B^{(*)}\bar{B}^{(*)}\to
\Upsilon(mS)\pi^+\pi^-$, and may change the distribution of the
$\Upsilon(mS)\pi^+\pi^-$ signals significantly around the peak of
$\Upsilon(5S)$ resonance.

In general, neglecting radiative corrections and the beam-energy
spread, the cross section for the process $e^+e^-\to$ resonance $r$
$\to$ hadronic final state $f$ at the center-of-mass energy
$\sqrt{s}$ can be approximately expressed by the Breit-Wigner form
\be \sigma(s)=12\pi\frac{\Gamma^{ee}_r
\cdot\Gamma^{f}_r(s)}{(s-m_r^2)^2+m_r^2\Gamma_r(s)^2},\label{sigma(s)}\ee
where the resonance $r$ is parameterized by its mass $m_r$, the
total width $\Gamma_r$, the electronic width $\Gamma^{ee}_r$, and
the partial width $\Gamma^{f}_r$ for the decay channel $f$. For the
$\Upsilon(5S)$, in (\ref{sigma(s)}) we have neglected the weak
$s$-dependence of $m_r$, $\Gamma^{ee}_r$. However, for the final
states $f=\Upsilon(1S/2S/3S)\pi^+\pi^-$, the $s$-dependence of
$\Gamma^{f}_r$ is as strong as $|\vec{p}_1(s)|^6$, and therefore can
not be neglected at all.

\begin{figure}[t]
\begin{center}
\vspace{0cm}
 \hspace*{0cm}
\scalebox{0.5}{\includegraphics[width=15cm,height=11cm]{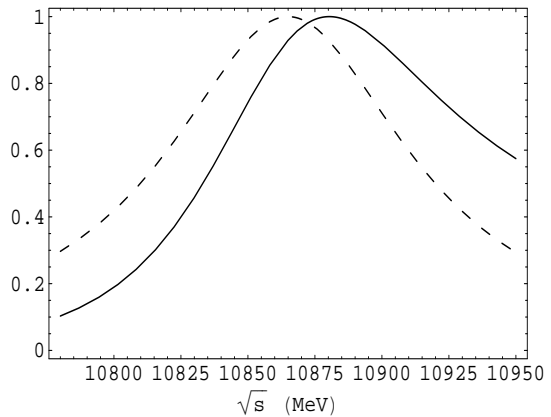}}
\end{center}
\vspace{0cm}\caption{Resonance line-shapes of the naive Breit-Wigner
distribution (dashed line) and the one with
$\Gamma^{f}_r(s)\sim|\vec{p}_1(s)|^6$ and $B\bar{B}$ as the
intermediate state (solid line) . The maximums are normalized to
one, respectively.}\label{Fig:BW-pCM6}
\end{figure}

As a first step, we will neglect the energy dependence of the total
width of $\Upsilon(5S)$ as well. Thus, for the final state, such as
$f=light~hadrons$, the energy distribution of the cross section of
$\Upsilon(5S)$ will peak at $m_{5S}$ exactly as that shown in
Fig.~\ref{Fig:BW-pCM6} by the dashed line. We call this the naive
Breit-Wigner distribution. In Fig.~\ref{Fig:BW-pCM6}, the resonance
parameters of $\Upsilon(5S)$ have been chosen from
PDG2006~\cite{PDG06} with the following central values of $m_{5S}$
and $\Gamma_{5S}$:
\be
m_{5S}=10865~\mbox{MeV},~~~~~~~\Gamma_{5S}=110~\mbox{MeV}\label{m5S-Gamma5S}.\ee
Then, we can compare the naive Breit-Wigner distribution with the
one where $\Gamma^{f}_{5S}(s)\sim|\vec{p}_1(s)|^6$. For the
$B\bar{B}$ intermediate state, the result is shown in
Fig.~\ref{Fig:BW-pCM6} with the solid line. One can see that the
strong $s$-dependence of $\Gamma^{f}_{5S}$ pushes the resonance peak
up with a energy shift of about 15 MeV.


The two-body decay modes $B^{(*)}\bar{B}^{(*)}$ and
$B^{(*)}_{s}\bar{B}^{(*)}_{s}$, which are dominant ones of
$\Upsilon(5S)$~\cite{PDG06}, will bring on large energy dependence
of $\Gamma_{5S}$ and change the distributions in
Fig.~\ref{Fig:BW-pCM6} consequently. For simplicity, we will choose
the main decay mode $B^*\bar{B}^*$ to estimate the energy dependence
of $\Gamma_{5S}$. That is,
\be
\Gamma_{5S}(s)\sim\Gamma^{B^*\bar{B}^*}_{5S}(s)\sim\frac{s+3m_{B^*}^2}{s}\Big(\frac{\sqrt{s-4m_{B^*}^2}}{2}\Big)^3,\label{Gamma5S:s-dependence}\ee
which can be derived from the effective Lagrangian (\ref{L-YB*B*})
with $g_{\Upsilon(5S)B^*B^*}$ being treated as a constant
independent of $s$. Here, $\Gamma_{5S}(m_{5S})$ is normalized to be
110 MeV, like that in (\ref{m5S-Gamma5S}).

\begin{figure}[t]
\begin{center}
\vspace{0cm}
 \hspace*{0cm}
\scalebox{0.5}{\includegraphics[width=15cm,height=11cm]{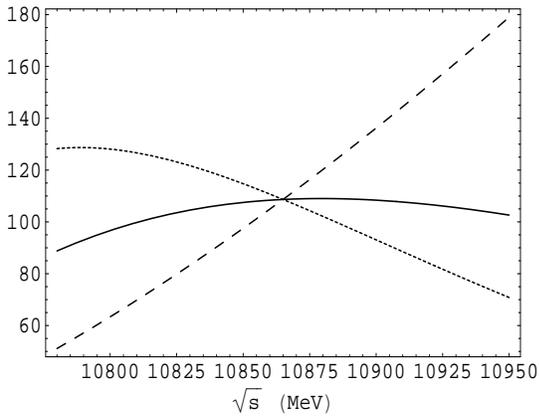}}
\end{center}
\vspace{0cm}\caption{The energy dependence and $\alpha$ dependence
of $\Gamma_{5S}(s)$. The dashed line, solid line and dotted line are
evaluated with $\alpha=0.0$, 0.6 and 1.0 GeV$^{-2}$,
respectively.}\label{Fig:Gam5S-s}
\end{figure}

Furthermore, one need introduce form factors for every
$\Upsilon(5S)B^{(*)}\bar{B}^{(*)}$ vertexes in Fig.~\ref{Fig:Y-YS}.
Following Ref.~\cite{Pennington}, we choose the form factor
\be
\mathcal{F}(s)=\frac{\mbox{Exp}(-\alpha|\vec{p}_1(s)|^2)}{\mbox{Exp}(-\alpha|\vec{p}_1(m^2_{5S})|^2)}\,,\label{FormFactor}
\ee
which is normalized to 1 at $s=m_{5S}^2$, since we have extracted
the on-shell coupling constants $g_{\Upsilon(5S)B^{(*)}B^{(*)}}$ in
(\ref{g:Upsilon(5S)BB}-\ref{g:Upsilon(5S)B*B*}) at this energy
point. This form factor, which is associated with the effective
$\Upsilon(5S)B^{(*)}\bar{B}^{(*)}$ coupling, can be understood as
the consequence of the overlap integral of the wave functions of
$\Upsilon(5S)$ and the two $B^{(*)}$ mesons, and the scale factor
$\alpha$ can be related to the effective radius of the interaction,
$R$, by $\alpha= R^2/6$~\cite{Pennington}. It is this form factor
(or others of a similar form) that plays the role to naturally
balance the otherwise over-increased decay rates with increased
phase space. Namely, this form factor is a sort of "cutoff" for the
high momentum of the final state mesons. The momentum dependence of
the effective couplings can partially explain why the coupling
constants $g_{\Upsilon(5S)B^{(*)}B^{(*)}}$ are smaller than the one
$g_{\Upsilon(4S)BB}$, though for $\alpha=0\mbox{-}1$ GeV$^{-2}$ the
suppression from the form factor is not strong enough to account for
the large difference between the values in (\ref{g:Upsilon(4S)BB})
and in (\ref{g:Upsilon(5S)BB}-\ref{g:Upsilon(5S)B*B*}). This
suppression may be additionally due to the node structure of the
wave function of highly excited $\Upsilon(5S)$~\cite{Meng-Y5S}.

\begin{figure}[t]
\begin{center}
\vspace{0cm}
 \hspace*{0cm}
\scalebox{0.5}{\includegraphics[width=15cm,height=11cm]{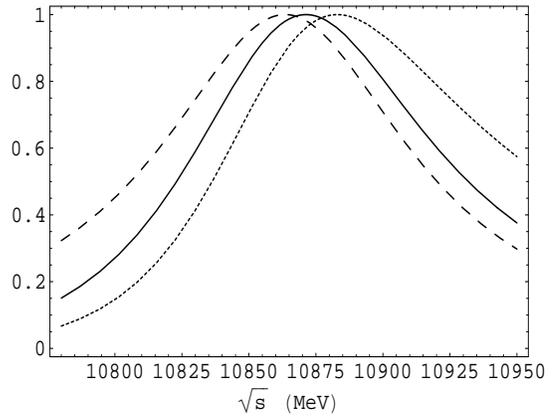}}
\end{center}
\vspace{0cm}\caption{Resonance line-shapes  for $f=B^*\bar{B}^*$
(dashed line) and the one for $f=\Upsilon(1S)\pi^+\pi^-$ (solid
line) and for $f=\Upsilon(3S)\pi^+\pi^-$ (dotted line) with
$\alpha=0.6$ GeV$^{-2}$. The maximums are normalized to one,
respectively.}\label{Fig:BW-1S-3S}
\end{figure}

Needless to say, the form factor in (\ref{FormFactor}) should also
change the energy dependence of the width
$\Gamma_{5S}^{B^{(*)}\bar{B}^{(*)}}(s)$ and the total width
$\Gamma_{5S}(s)$ in (\ref{Gamma5S:s-dependence}) subsequently. Such
$\alpha$ dependence of $\Gamma_{5S}(s)$ is shown in
Fig.~\ref{Fig:Gam5S-s}. One can see that the energy dependence is
significantly weakened when $\alpha$ is about 0.6 GeV$^{-2}$ (solid
line in Fig.~\ref{Fig:Gam5S-s}). As a result, the distribution of
$\Upsilon(5S)$ with $f=B^*\bar{B}^*$ and $\alpha=0.6$ GeV$^{-2}$,
which is shown in Fig.~\ref{Fig:BW-1S-3S} with the dashed line, is
very close to the naive Breit-Wigner one.

With $\alpha=0.6$ GeV$^{-2}$~\cite{Pennington}, the distributions
for $f=\Upsilon(1S)\pi^+\pi^-$ can be evaluated numerically. Since
here we only focus on the line-shape and the peak shift, which are
almost independent of the parameters except $\alpha$ and
$\Gamma_{5S}(m_{5S})$, it is quite safe to choose the central values
for other parameters introduced in the last section. The
distribution for $f=\Upsilon(1S)\pi^+\pi^-$ is illustrated in
Fig.~\ref{Fig:BW-1S-3S} with the solid line. Compared with the
distribution for $f=B^*\bar{B}^*$ (the dashed line), the peak shift
is about 7 MeV. If we choose the channel $B\bar{B}$ as the main
hadronic channel, the peak shift can be as large as 11 MeV for the
$\Upsilon(1S)\pi^+\pi^-$ channel.

The energy distribution for $f=\Upsilon(2S)\pi^+\pi^-$ has similar
line-shape to that for $f=\Upsilon(1S)\pi^+\pi^-$ within the energy
region of 10780-10950 MeV.  Furthermore, since the transition
$\Upsilon(5S)\to\Upsilon(3S)\pi^+\pi^-$ is also
observed~\cite{Belle07-5SmS} with a large rate and quite high
statistical significance ($3.2\sigma$), we also evaluate its energy
distribution in Fig.~\ref{Fig:BW-1S-3S} with the dotted line. We
find there is an additional peak shift of about 12 MeV relative to
the distribution for $f=\Upsilon(1S)\pi^+\pi^-$. This is just
because the role of the scalar resonance in Fig.~\ref{Fig:Y-YS}. In
our model~\cite{Meng-Y5S}, the final-state $\pi^+\pi^-$ are assumed
to be dominated by the scalar resonance
$\mathcal{S}$($\sigma,f_0(980)$...)
~\cite{Komada01-scalar-contributions}, which can be described by the
Breit-Wigner distribution (\ref{BW:scalar}) at the cross-section
level. If the mass of the $\sigma$ resonance is chosen as
$m_{\sigma}=526$ MeV given in Tab.~\ref{Tab:Scalar-resonance}, then
one can easily find that the mass difference
$m_{5S}-m_{3S}\approx510$ MeV happens to be in the center of the
distribution in (\ref{BW:scalar}). As we have mentioned in the end
of the last section, this brings on another strong $s$-dependence to
the cross section for $f=\Upsilon(3S)\sigma(\pi^+\pi^-)$. Thus, the
distribution for $f=\Upsilon(3S)\pi^+\pi^-$ is pushed up farther
than those for $f=\Upsilon(1S,2S)\pi^+\pi^-$. Similarly, there will
be a long tail in the distribution for $f=\Upsilon(2S)\pi^+\pi^-$
within the energy region of 10950-11100 MeV due to the emergence of
the scalar resonance $f_0(980)$.

It is worth emphasizing that the scalar resonance dominance is just
a simplification to study the effects of the final state
rescattering mechanism in the
$\Upsilon(5S)\to\Upsilon(1S,2S,3S)\pi^+\pi^-$
transitions~\cite{Meng-Y5S}. Thus, the differences between the
distributions for $f=\Upsilon(1S,2S,3S)\pi^+\pi^-$ tend to disappear
if the non-resonance contributions and their interference with the
scalar resonances are important. Therefore, to detect the
differences between the distributions for
$f=\Upsilon(1S,2S,3S)\pi^+\pi^-$ can provide useful information on
the role of the scalar resonances in these transitions. More
valuable information can also be given by the measurements on the
$M(\pi\pi)$ spectrum and the distribution of the helicity
angle~\cite{Belle07-5SmS}.

\begin{figure}[t]
\begin{center}
\vspace{0cm}
 \hspace*{0cm}
\scalebox{0.5}{\includegraphics[width=16cm,height=6.8cm]{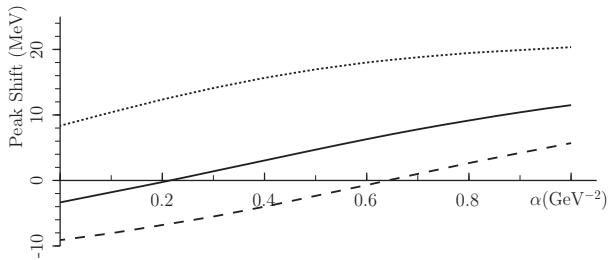}}
\end{center}
\vspace{0cm}\caption{$\alpha$-dependence of the peak shifts compared
with $m_{5S}=10865$ MeV for $f=B^*\bar{B}^*$ (dashed line),
$f=\Upsilon(1S)\pi^+\pi^-$ (solid line) and
$f=\Upsilon(3S)\pi^+\pi^-$ (dotted
line).}\label{Fig:PeakShift-alpha}
\end{figure}

It is obvious that the position of the resonance peaks of
$\Upsilon(5S)$ in $\Upsilon(1S,2S,3S)\pi^+\pi^-$ channels depend on
the value of the scale factor $\alpha$ in the form factor
(\ref{FormFactor}). However, the energy distributions of
$\Upsilon(5S)$ in other main hadronic decay channels (e.g.,
$B\bar{B}$...) also depend on the same factor $\alpha$. So, it is
essential to compare the line-shape for e.g.
$f=\Upsilon(1S)\pi^+\pi^-$ with those main hadronic decay channels
with different choices of the value of the scale factor. For
$\alpha=0.0\mbox{-}1.0$ GeV$^{-2}$, the results are shown in
Fig.~\ref{Fig:PeakShift-alpha}, where the $B^*\bar{B^*}$ channel is
used to serve as one of the main hadronic channels again.

In Fig.~\ref{Fig:PeakShift-alpha},  with a reasonable choice for the
cutoff factor $\alpha$, say, 0.3-0.8 GeV$^{-2}$, the relative peak
shifts for $f=\Upsilon(1S,3S)\pi^+\pi^-$ are found to be almost
independent of the cutoff. Thus, we incline to conclude that there
should be a peak shift of about 7 MeV for the distribution of
$\Upsilon(5S)$ in the $\Upsilon(1S)\pi^+\pi^-$ channel compared with
that in the $B^*\bar{B}^*$ channel. Similar shift should be obtained
in the $\Upsilon(2S)\pi^+\pi^-$ channel. On the other hand, the
shift in the $\Upsilon(3S)\pi^+\pi^-$ channel could be 12 MeV larger
than those in the $\Upsilon(1S,2S)\pi^+\pi^-$ channels, but this
depends on the assumption that the scalar resonance $\sigma$
dominates the $\Upsilon(3S)\pi^+\pi^-$ transition. Furthermore, if
the  $B\bar{B}$ is chosen as the main hadronic channel, the peak
shifts will be enlarged by $3\mbox{-}4$ MeV.

\section{Summary}

In summary, we study the energy distributions of dipion transitions
$\Upsilon(5S)$ to $\Upsilon(1S,2S,3S)\pi^+\pi^-$ in the final state
rescattering model. Since the $\Upsilon(5S)$ is well above the open
bottom thresholds, the dipion transitions can proceed through the
real process $\Upsilon(5S)\to B^{(*)}\bar{B}^{(*)}$ and the
subsequent process $B^{(*)}\bar{B}^{(*)}\to
\Upsilon(1S,2S,3S)\pi^+\pi^-$. This model can not only explain the
observed unusually large rates of $\Upsilon(5S)\to
\Upsilon(1S,2S,3S)\pi^+\pi^-$\cite{Meng-Y5S}, but also predict a
unique energy dependence of the cross sections of
$e^+e^-\to\Upsilon(5S)\to \Upsilon(1S,2S,3S)\pi^+\pi^-$. We find
that the energy distributions of $\Upsilon(1S,2S,3S)\pi^+\pi^-$
markedly differ from that of $\Upsilon(5S)\to B^{(*)}\bar{B}^{(*)}$,
and in particular, the resonance peak will be pushed up by about
7-20 MeV for these dipion transitions relative to the main hadronic
decay modes. These predictions can be used to test the final state
rescattering mechanism in hadronic transitions for heavy quarkonia
above the open flavor thresholds.


\begin{acknowledgments}
We are grateful to K.F. Chen, Y. Sakai, C.Z. Yuan, Z.P. Zhang, and
S.L. Zhu for useful discussions. One of us (K.T.C.) would like to
thank the Belle Collaboration and Y. Sakai for the hospitality
during his visit to KEK, Japan. This work was supported in part by
the National Natural Science Foundation of China (No 10675003, No
10721063).
\end{acknowledgments}

\bibliography{apssamp}

\begin{thebibliography}{99}

\bibitem{Kuang06} Y.~P.~Kuang, Front.\ Phys.\ China {\bf 1}, 19
(2006); M.B. Voloshin, arXiv:0711.4556[hep-ph], to appear in Prog.
Part. Nucl. Phys.; E. Eichten, S. Godfrey, H. Mahlke, J.L. Rosner,
arXiv:hep-ph/0701208.

\bibitem{Olsen}
S.L.~Olsen, arXiv: 0801.1153 [hep-ex], to appear in Chin. Phys. C.

\bibitem{hou} W.S. Hou, Phys.Rev. D74, 017504 (2006).

\bibitem{Belle07-5SmS} K.F. Chen et al. [Belle
Collaboraion], arXiv: 0710.2577 [hep-ex].

\bibitem{Meng-Y5S}
C. Meng, K.T. Chao,  Phys. Rev. D77, 074003 (2008) (arXiv: 0712.3595
[hep-ph]).

\bibitem{Cheng05_ReSC}H.Y. Cheng, C.K. Chua and A. Soni,
Phys. Rev. D71, 014030 (2005); P. Colangelo, F. De Fazio and T. N.
Pham, Phys. Lett. B542 71 (2002); Phys. Rev. D69 054023 (2004); X.
Liu, B. Zhang and S.L. Zhu, Phys. Lett. B645, 185 (2007).

\bibitem{Simonov07-nSmS} Yu.A.~Simonov, JETP Lett. 87 147 (2008) (arXiv: 0712.2197 [hep-ph]); arXiv: 0804.4635 [hep-ph].

\bibitem{Meng07_X3872_ReSC}
C. Meng, K.T. Chao, Phys. Rev. {\bf D75}, 114002 (2007).
\bibitem{Meng07_Z4430_ReSC}
C. Meng, K.T. Chao, arXiv:0708.4222[hep-ph].

\bibitem{PDG06}
W.~M. Yao et al., J. Phys. G {\bf 33}, 1 (2006); see also the
Website: http://pdg.lbl.gov/.



\bibitem{Komada01-scalar-contributions}
T.~Komada, M.~Ishida and S.~Ishida, Phys. Lett. B {\bf 508}, 31
(2001); {\bf 518}, 47 (2001).

\bibitem{Pennington}
M.R.~Pennington and D.J.~Wilson, Phys. Rev. {\bf D76}, 077502
(2007).







\end{thebibliography}

\end{document}